**Ya. D. Krivenko-Emetov\*, O. S. Shevchuk**

*National Technical University of Ukraine «Igor Sikorsky Kyiv Polytechnic Institute»*, *Kyiv, Ukraine*

\*Corresponding author: y.kryvenko-emetov@kpi.ua; krivemet@ukr.net

# INVESTIGATION OF THE INTERNAL STRUCTURE OF THE DEUTERON AGAINST THE BACKGROUND OF TWO-PHOTON EXCHANGE EFFECTS IN ELASTIC ELECTRON-DEUTERON SCATTERING[1]

The aim of this work is to investigate the simultaneous influence of two-photon effects in quantum electrodynamics (QED) and logarithmic corrections in quantum chromodynamics (QCD) on certain observable experimental quantities (structure functions $A(Q^2)$ and $B(Q^2)$ in elastic electron-deuteron scattering. Analyzing these effects broadens our understanding of electron scattering physics, particularly the manifestations of quark-gluon degrees of freedom in the deuteron.

## 1. Introduction

To date, a vast amount of experimental data on the interaction of polarized and unpolarized deuterons with electrons at large values of the transferred momentum squared, $Q^2$, has been accumulated. This opens up new opportunities for studying the structure of the deuteron at distances smaller than the nucleon size. However, the comparison of predictions from perturbative Quantum Chromodynamics (pQCD), obtained from the analysis of lower-order quark-quark interactions, with available experimental data and other approaches to elastic scattering of electrons by polarized and unpolarized deuterons remains insufficiently explored. In the asymptotic region, where the magnitude $Q^2$ significantly exceeds the deuteron's mass squared, according to pQCD, predictions for the functional dependence of the deuteron form factors can be obtained based on the phenomena of asymptotic freedom and the factorization theorem. In this case, the deuteron is considered a system of 6 quarks moving collinearly, each contributing to the deuteron's momentum fraction. However, these pQCD predictions can be conditionally divided into two parts: predictions based on the "quark counting rules" (the so-called "cascade" pQCD diagrams (Fig. 1, a), which are relatively well-supported by comparisons with experimental data [1 – 6]), and more subtle corrections which, although derived from the analysis of dominant contributions from simple ladder-type pQCD diagrams ("quark exchange" diagrams (Fig. 1, b)) [7 - 11], do not have such consistent experimental confirmation and, therefore, have not gained as wide acceptance as the predictions based on the "quark counting rules" [5, 6, 12].

---

[1] Presented at the XXXI Annual Scientific Conference of the Institute for Nuclear Research of the National Academy of Sciences of Ukraine, Kyiv, May 27 - 31, 2024.

On the other hand, in recent decades, the role of higher-order perturbation theory beyond the single-photon approximation in electron scattering on hadronic systems has been widely discussed within the framework of QED. This is mainly associated with precise measurements of the electric and magnetic form factors of the proton conducted at the Thomas Jefferson National Accelerator Facility (JLab) [13, 14] and their theoretical interpretation [15 - 17].

As a result of the aforementioned studies, it has been shown that accounting for two-photon contributions leads to qualitative changes in both the differential cross-section and polarization observables [15 - 20]. However, considering two-photon contributions and logarithmic pQCD corrections separately has not yet led to a significant breakthrough in the experimental description of elastic electron-deuteron scattering [5, 11, 21] in the context of comparing theory and data.

This article is dedicated to comparing the combined contribution of these two scattering mechanisms with experimental data [6, 22 - 24] (similar to how it was done in [11] and [21]).

## 2. A brief theoretical description and problem statement

In the pQCD approach, at high energies, the masses of quarks and hadrons are neglected. In this case, the amplitude of the investigated process is expressed through the amplitude of hard electron-quark scattering, multiplied by the nonperturbative part (which can be associated with parameters $N_i$ ($i=1,2,3$)[2], which is associated with the distribution functions of quarks and gluons in the deuteron in the initial and final states [1 - 4, 7, 12]. When calculating the amplitude of hard (specifically perturbative) scattering, the deuteron is considered as a system of 6 quarks moving collinearly, each of which contributes to the deuteron's momentum fraction: $x_i = p_i^+/P^+$, where $p_i^+ = p_i^0 + p_i^3$, $0 < x_i < 1$, $\sum_i x_i = 1$ (given in the light-cone formalism).

In Fig. 1, a, b schematic diagrams of elastic electron-deuteron scattering are presented (the horizontal lines represent quarks, the curved lines between the quarks are gluons, the propagator of each gluon contributes $\alpha_s(Q^2)/Q^2$ (where $Q^2 = -q^2$) to the hard amplitude of the process) $\alpha_s(Q^2)$ is the running strong coupling constant.

---

[2] $N_i$ are fitting parameters for determining the deuteron structure functions $A(Q^2)$ and $B(Q^2)$ (see Eq. (1) and Figs. 4 and 5)

The first "cascade" (Fig. 1, a) diagram corresponds only to the high-energy case of "hard" scattering, when, according to quark counting rules, the deuteron is represented by the minimal system of interacting valence quarks (with the contribution of sea quarks in the high-energy region being neglected). The second diagram (Fig. 1, b) accounts for more subtle exchange effects between two quark "clusters" from which, at lower energies, a neutron and a proton are formed. The transition from the 6-quark system to two clusters, each containing 3 quarks, is schematically shown in Fig. 2.

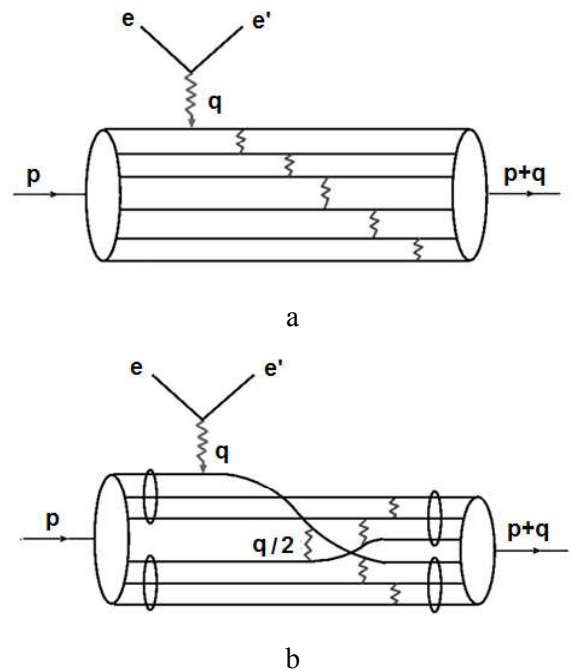

a

b

Fig. 1. The diagram based on the "quark counting rules" (a),
and the diagram based on "quark rescatterings" (b), leading to finer logarithmic corrections (the figure is taken from [5])

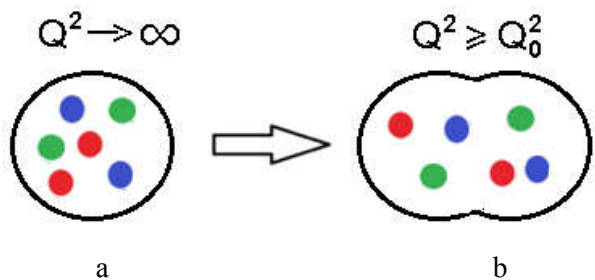

a          b

Fig. 2. The transition from the 6-quark system (a),
two quark clusters (each cluster "contains" three quarks) (b)

It is assumed that the main form factor of the deuteron in the perturbative region can be represented as the product of three factors, exhibiting dipole $G(Q^2/4) = 1/(1 + Q^2/(4\mu^2))^2$ [11], power-law, and logarithmic behavior, respectively [4, 7, 10, 11]:

$$G_1^d(Q^2) \sim N_1 G^2\left(\frac{Q^2}{4}\right)\left[\frac{\alpha_s^5(Q^2)}{Q^2}\right]\left(\frac{\ln^{-2\gamma^d}(Q^2/\Lambda_{QCD}^2)}{\ln^{-4\gamma^N}(Q^2/4\Lambda_{QCD}^2)}\right), \qquad (1)$$

where $N_1$ and $\mu$ are non-perturbative parameters, $\gamma^d = 6C_F/5\beta$, $C_F = (n_c^2 - 1)/(2n_c^2)$, $n_c = 3$ is a number of quark colors, $\beta = 11 - \frac{2}{3}n_f$, $n_f = 5$ is a number of quark flavors ( since the probability of the production of the sixth top quark in the energy range under study can be neglected), $\gamma^N = C_F/2\beta$ are so-called the "anomalous dimensions" of the deuteron and nucleon, depending on the number of flavors and colors, and $\Lambda_{QCD} = 0.2\,\text{GeV}/c$ is the characteristic scale factor of the pQCD. The parameter $\mu$ differs from the value $0.71\,\text{GeV}/c$ (as in [11]) for a free nucleon and thus takes into account the influence of the nuclear environment.

In the experimental data, two kinematic regions can be conventionally distinguished, separated by a characteristic stitching parameter $Q_0^2$. The first region is the "pQCD region" with relatively large values of the momentum transfer squared $Q^2 \geq Q_0^2$. TN_1he second region is the low-energy "meson region" ($Q^2 < Q_0^2$). The high-energy region is parameterized according to pQCD predictions [4, 7, 11, 12]. A "meson" approximation has been proposed for describing experimental data in the low-energy region [11, 12]. The parameter values $Q_0^2 = 3.5\,(\text{GeV}/c)^2$ [11] and $Q_0^2 = 9 - 40\,(\text{GeV}/c)^2$ [12], as well as the validity of the adopted parameterizations for both the low- and high-energy regions, were obtained from the analysis of experimental data.

In this work, since we are primarily interested in the high-energy region of pQCD and QED predictions, we do not perform stitching between the high- and low-energy regions. However, we implicitly rely on the results from [11] and [12], which provide estimates of the energy scales where quark degrees of freedom and higher-order two-photon contributions from QED perturbation theory begin to manifest. Based on the data from these studies, we limited the experimental data used from [6, 22 - 24] to a boundary of $Q_0^2 = 1.75\,(\text{GeV}/c)^2$.

However, there is no experimental data with sufficient statistics to assess the contribution of the exchange-type diagrams (see Fig. 1, b). Therefore, they have not received as wide recognition [5, 11]. On the other hand, within the framework of QED, the role of higher-order perturbation theory beyond the single-photon approximation in electron scattering on hadronic systems has been widely discussed [13 - 19].

Among second-order perturbation diagrams, the two-photon exchange diagram plays a particularly important role. Indeed, it is the only one with a structure different from the Born approximation, and therefore it leads to a qualitative change in the scattering amplitude structure.

In the two-photon approach, two types of TPE amplitudes are calculated: one associated with Feynman diagrams in which two photons interact with the same nucleon $M_2^I$ (Fig. 3, a) [16- 18], and the other with diagrams in which each of the two virtual photons interacts with different nucleons $M_2^{II}$ (Fig. 3, b) [19]. Based on the analysis of the graphs in [21] and the analytical formulas in [25], it can be assumed that TPE contributions can be effectively approximated by logarithmic functional dependencies.

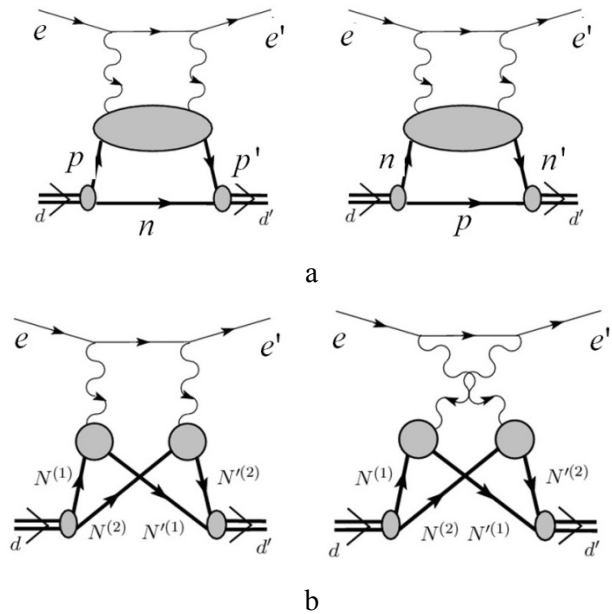

Fig. 3. Diagrams of two-photon exchange:

at amplitude $M_2^I = M_p^I + M_n^I$ (a), at amplitude $M_2^{II} = M_p^{II} + M_n^{II}$ (b).

## 3. Model description and obtained results

As noted above, accounting for higher orders of perturbation theory in pQCD [11] and QED [21] separately has not led to a significant improvement in describing experimental data [6, 22 - 24] (similar to how it was done in [11] and [21]). At the same time, considering the importance of these effects, it would be interesting to see how the degrees of freedom of quarks and gluons manifest in elastic electron-deuteron scattering against the background of two-photon contributions. For example, this can be done by comparing the theoretical description of experimental data on elastic

$e-d$ scattering with and without consideration of two-photon corrections. It would be interesting to choose the most advanced theoretical model – namely, the model that takes into account the logarithmic corrections of pQCD extending into the "meson" low-energy region, as the baseline theoretical model.

Based on the theoretical and experimental analysis conducted previously [19] [21] and [25] beyond the single-photon order of perturbation theory ($A_{1ph}(Q^2)$ and $B_{1ph}(Q^2)$ are the target and magnetic structure functions in the one-photon approximation), we propose the following phenomenological parameterization of two-photon corrections (written in terms of the fine structure constant in natural units, $\alpha = e^2/4\pi = 1/137$, $\lambda_{EM}^2 = Q_0^2 = 1.75\,(\text{GeV}/c)^2$):

$$A(Q^2) = A_{1ph}\left(1 + a_A \times \alpha \ln^{b_A}\left(Q^2/\lambda_{EM}^2\right)\right), \tag{2}$$

$$B(Q^2) = B_{1ph}\left(1 + a_B \times \alpha \ln^{b_B}\left(Q^2/\lambda_{EM}^2\right)\right), \tag{3}$$

where $a_A$, $a_B$ and $b_A$, $b_B$, are fitting parameters of the model, the structure functions $A(Q^2)$ and $B(Q^2)$ are expressed through

$$\frac{d\sigma}{d\Omega} = \left(\frac{d\sigma}{d\Omega}\right)\bigg|_{Mott} \left\{ A(Q^2) + B(Q^2) \tan^2 \frac{\theta_{Lab}}{2} \right\}$$

that is the differential cross section for unpolarized particles at the scattering angle $\theta_{Lab}$ in the laboratory coordinate system,

$$\left(\frac{d\sigma}{d\Omega}\right)\bigg|_{Mott} = \frac{\alpha^2}{4E_e^2} \frac{\cos^2\frac{\theta_{Lab}}{2}}{\sin^4\frac{\theta_{Lab}}{2}} \frac{1}{1 + \frac{2E_e^2}{M}\sin^2\frac{\theta_{Lab}}{2}} \text{ is the Mott cross section,}$$

$E_e$ is the energy of the initial electron,

$$A(Q^2) = G_C^2 + \frac{2}{3}\eta G_m^2 + \frac{8}{9}\eta^2 G_Q^2, \quad B(Q^2) = \frac{4}{3}\eta(1+\eta)G_M^2 \text{ are the target structure function and the}$$

magnetic structure function, respectively,

$G_C, G_M, G_Q$, which are functions of $G_1$, $G_2$, and $G_3$ (for more details, see, for example, [11]), are called the charge, magnetic, and quadrupole form factors of the deuteron, respectively, and $\eta = Q^2/(4M^2)$, where $M$ is the mass of the deuteron.

The exact formulas obtained in works [19] and [21] are difficult to parameterize using simple functions due to their analytical complexity. However, based on the analysis of the graphs showing the dependence of $A(Q^2)$ and $B(Q^2)$, obtained in [21], we can hypothesize (this is the main assumption of our work) that the contribution of two-photon corrections can potentially be effectively described by a logarithmic approximation (the logarithmic behavior of the two-photon contributions is most clearly visible in [21] for polarization tensor component, $T_{22}$). We understand that this assumption has not yet been strictly proven, but since this work presents a phenomenological analysis, we hope that it may be of interest as the first attempt to simultaneously account for exchange pQCD corrections alongside two-photon corrections. Moreover, in [25], analytical behavior was obtained when analyzing TPE effects for ep-scattering.

In this study, the pQCD predictions (1) are compared with experimental data for the structure functions, taking into account two-photon corrections according to parameterizations (2), (3), and without considering these corrections.

The figures below show the comparison of the asymptotic "logarithmic" prediction of pQCD with the proposed parameterization of two-photon corrections against the experiment (Figs. 4 and 5, solid thick line). For comparison, the same logarithmic pQCD prediction (1), but without accounting for two-photon corrections, is presented on the same graphs (Figs. 4 and 5, dashed thin line). It has been found that the asymptotic logarithmic behavior predicted by pQCD slightly improves in the presence of two-photon corrections compared to without them (the chi-square value $\chi^2_{2ph} = 14.160$ (and $\chi^2/NDF$ per number of degree of freedom (NDF): 1.09)[3], considering two-photon contributions is lower than the chi-square value $\chi^2_{1ph} = 152.658$ ($\chi^2/NDF$ per number of degree of freedom: 8.98) obtained without considering these contributions). For the red solid line, the obtained value is $\chi^2_A = 10.87$ (and $\chi^2_A/NDF = 1.20$ per degree of freedom); for the red dashed line, the obtained value is $\chi^2_A = 138.531$ (and the corresponding $\chi^2_A/NDF = 12.594$ per degree of freedom); for the blue solid line, the obtained value is $\chi^2_B = 3.29$ (and the corresponding $\chi^2_B/NDF = 3.29$ per degree of freedom); for the blue dashed line, the obtained value is $\chi^2_B = 14.127$ (and the corresponding $\chi^2_B/NDF = 4.709$ per degree of freedom).

The model uses the most reliable high-energy experimental points based on a critical analysis of known experimental data [6, 22 - 24]. The comparison has been conducted for models with 7 (out of which there are 3 common parameters, that is, 5 parameters for $A(Q^2)$ and 5 for $B(Q^2)$ separately) and 3 independent fitting parameters, respectively, as well as with one common parameter, which we fixed in both cases with the value $n_f = 5$. A sample of 20 points has been taken, and a fitting criterion $\chi^2$ has been obtained.

It is interesting to compare these results with the results of other authors [6, 11 - 12, 20]. It's important to note the key observation from [11] to justify the proposed basic model with logarithmic pQCD corrections. In the article [12], where anomalous dimensions were not taken into account, the description of the data was less accurate. However, in the paper [12], a larger number of fitting parameters were used, namely 13, compared to 10 in paper [11]. It can be argued that subsequent fittings also did not lead to significantly better results [5, 6]. Therefore, in our study, we used the parameterization that takes into account the logarithmic corrections of pQCD from [11] as the base, rather than the simple power-law parameterization from [12].

---

[3] For comparison, we also attempted to fit the same set of experimental points with a simple four-parameter power-law model for structure functions $A(Q^2)$ and $B(Q^2)$, which resulted in a significantly worse description of the experiment (overall chi-squared = 40.46 and the corresponding chi-squared per degree of freedom = 2.53).

The asymptotic logarithmic behavior predicted by the pQCD has been found to improve slightly with two-photon corrections (b) compared to without them (a) (respectively, solid and dotted lines in Figs. 4 and 5).

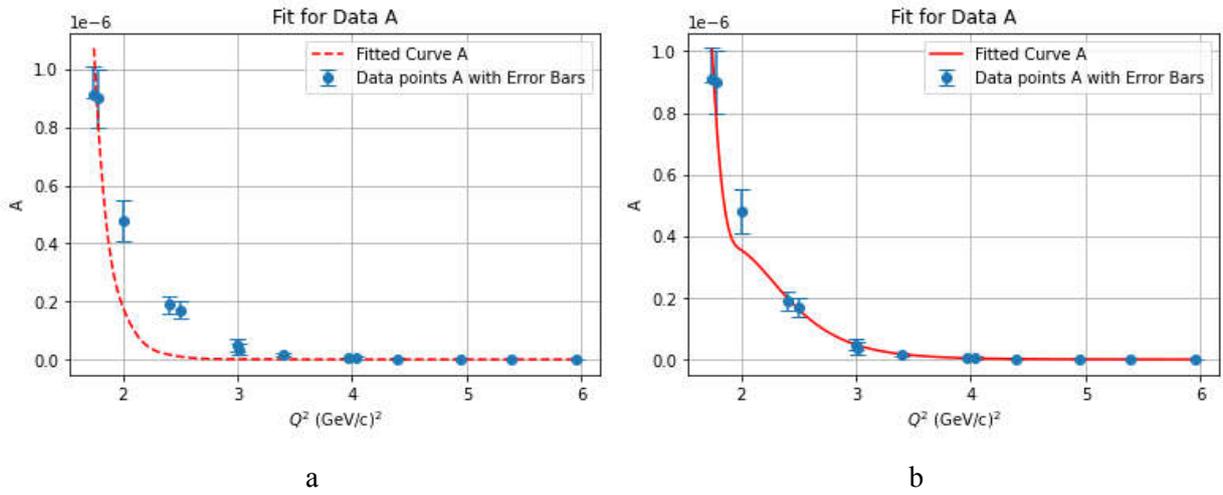

a                                                    b

Fig. 4. Optimal fits for the deuteron structure function $A(Q^2)$. The uncertainties on the data points are statistical and systematic uncertainties combined in quadrature. The data are taken from [6, 23, 24]

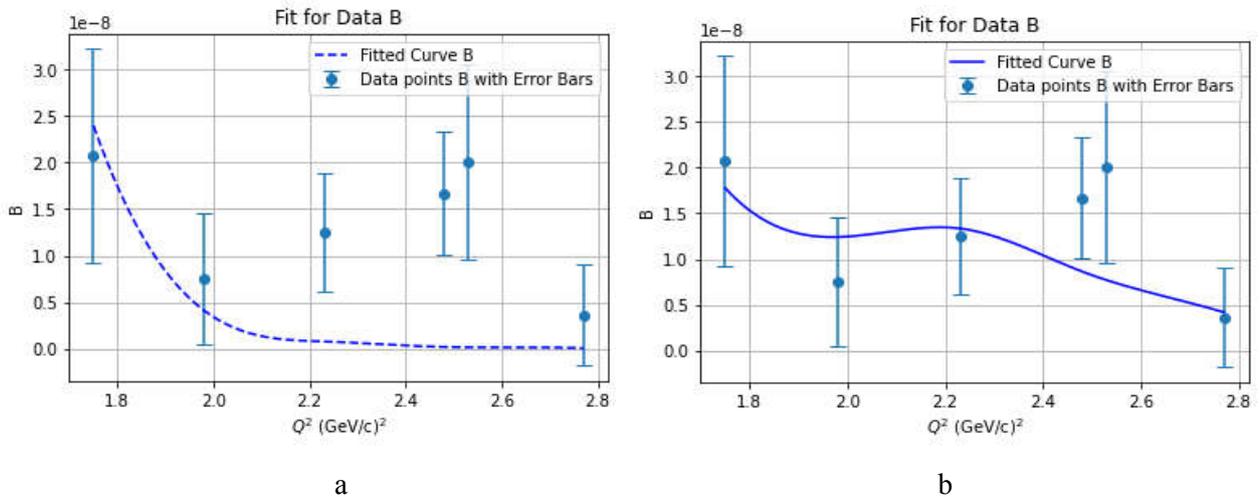

a                                                    b

Fig. 5. Optimal fits for the deuteron structure function $B(Q^2)$, designations as in Fig. 4. The uncertainties on the data points are statistical and systematic uncertainties combined in quadrature. The data are taken from [22]

In the figures, the solid line describes the experimental data by using the QCD logarithmic corrections + QED two-photon corrections, while the dash line describes the data by only using the QCD logarithmic corrections.

The optimal fitted parameters for $A(Q^2)$ are as follows: dotted line – $N_1 = 1.427 \cdot 10^{-1}$, $N_2 = 1.727 \cdot 10^{-2}$, $N_3 = 5.490 \cdot 10^{-3}$; solid line – $N_1 = 1.479 \cdot 10^{-1}$, $N_2 = 1.491 \cdot 10^{-2}$, $N_3 = 3.297 \cdot 10^{-2}$, $a_A \times \alpha = 287.545$, $b_A = 2.780$.

The optimal fitted parameters for $B(Q^2)$ are as follows: dotted line – $N_1 = 1.427 \cdot 10^{-1}$, $N_2 = 1.727 \cdot 10^{-2}$, $N_3 = 5.490 \cdot 10^{-3}$; solid line – $N_1 = 1.479 \cdot 10^{-1}$, $N_2 = 1.491 \cdot 10^{-2}$, $N_3 = 3.297 \cdot 10^{-2}$, $a_B \times \alpha = 1719.536$, $b_B = 2.99$.

From a physical standpoint, we restricted the values of the parameters $b_A$, and $b_B$ to be less than 3.

## Summary

Based on the analysis of the results from the papers [1 - 7, 9, 11, 12, 15-19, 21, 25], dedicated to studying the influence of two-photon contributions and logarithmic corrections in elastic electron-deuteron scattering, the following conclusions can be drawn:

1. The analysis of these papers allows us to conclude that both effects are important in studying the interaction of electrons and deuterons. Logarithmic QCD corrections, as another key aspect of the study, have shown their unique impact on examining the internal structure of the deuteron. Two-photon QED effects, arising from the interaction between electrons and deuterons, largely due to their interfering nature, demonstrate the importance of considering them to achieve high precision in describing experimental data.

2. Neither effect considered individually is sufficient to satisfactorily describe the existing experimental data.

Based on the conducted analysis, we have proposed a phenomenological model to account for both effects simultaneously (QCD logarithmic corrections + QED two-photon corrections) (Eqs. (2) and (3)).

Our study shows that logarithmic corrections are more significant than two-photon effects, due to a smaller number of parameters (3 independent fitting parameters compared to 7 parameters in the complex model). However, they substantially improve the fitting criterion only when combined with two-photon corrections (see Figs. 4 and 5).

The results of our comparison with experimental data highlight the importance of simultaneously accounting for two-photon effects and logarithmic corrections in the electron-deuteron interaction process, indicating the need to improve theoretical models to ensure consistency with experimental observations. They also point to the direction for further research in this area, namely carrying out experiments with larger statistics and expanding theoretical approaches to comprehensively account for all defining physical phenomena in elastic electron-deuteron scattering.

REFERENCES

1. G. P. Lepage, S. J. Brodsky. Exclusive Processes in Perturbative Quantum Chromodynamics. Phys. Rev. D 22 (1980) 2157.
2. V. A. Matveev, R. M. Muradyan, A. N. Tavkhelidze. Automodellism in the large-angle elastic scattering and structure of hadrons. Lett. Nuovo Cim. 7 15 (1973) 719-723.
3. S. J. Brodsky, R. F. Glennys. Scaling Laws at Large Transverse Momentum. Phys. Rev. Lett. 31 (1973) 1153–1156.
4. G. P. Lepage, S. J. Brodsky. Exclusive processes in perturbative quantum chromodynamics. Phys. Rev. D 22 (1980) 2157-2198.
5. Jingyi Zhou et al. Lowest-order QED radiative corrections in unpolarized elastic electron–deuteron scattering beyond the ultra-relativistic limit for the proposed deuteron charge radius measurement at Jefferson laboratory. Eur. Phys. J. A 59 (2023) 256.
6. D. Abbott et al. [JLab t20 Collaboration] Phenomenology of the deuteron electromagnetic form factors. European Physical Journal A 7 (2000) 421. [nucl-ex/0002003]. http://irfu.cea.fr/dphn/T20/Parametrisations/ and https://irfu.cea.fr/dphn/T20/WorldDataSet/index.php.
7. G. P. Lepage, S. J. Brodsky. Quantum chromodynamic predictions for the deuteron form facto Physical Review Letters 51 (2), (1983) 1-11.
8. A. V. Efremov, A. V. Radyushkin. Asymptotic behavior of the pion form factor in quantum chromodynamics. Theor. Math. Phys. 42 2 (1980) 97-110.
9. P. Ball, V. M. Braun. Higher twist distribution amplitudes of vector mesons in QCD: twist-4 distributions and meson mass corrections. Nucl. Phys. B 543 (1999) 201-238 [arXiv:hep-ph/9810475].
10. V. M. Braun et al. Baryon distribution amplitudes in QCD. Nucl. Phys. B 553 (1999) 355-426 [arXiv:hep-ph/9902375].
11. A. P. Kobushkin, Ya. D. Krivenko-Emetov. pQCD phenomenology of elastic ed scattering. Scientific works of the Institute of Nuclear Research. 3 (2003) 49–69 https://doi.org/10.48550/arXiv.nucl-th/0112009.
12. A. P. Kobushkin, A. I. Syamtomov. Deuteron Electromagnetic Form Factors in the Transitional Region Between Nucleon-Meson and Quark-Gluon Pictures. Physics of Atomic Nuclei 58 (1995) 1565–1571. DOI: https://doi.org/10.48550/arXiv.hep-ph/9409411.
13. M. K. Jones et al. GEp/GMp Ratio by Polarization Transfer in ep → e'p'. Phys. Rev. Lett. 84 (2000) 1398–1402.
14. O. Gayou et al. Measurement of GEp/GMp in e'p → ep' to Q2 = 5.6GeV$^2$. Phys. Rev. Lett. 88 (2002) 092301-092307.

15. A. I. Akhiezer, A. G. Sitenko. Diffractional Scattering of Fast Deuterons by Nuclei. Phys. Rev. 106 (1957) 1236-1246.
16. D. Borisyuk, A. Kobushkin. Box diagram in the elastic electron-proton scattering. Phys. Rev. C. 74 (2006) 065203(R)-065211(R).
17. D. Borisyuk, A. Kobushkin. Phenomenological analysis of two-photon exchange effects in proton form factor measurements. Phys. Rev. C. 76 (2007) 022201(R)- 022203(R).
18. Yu Bing Dong, D.Y. Chen. Two-photon exchange effect on deuteron electromagnetic form factors. Physics Letters B. 675 (2009) 426–432. DOI: https://doi.org/10.1016/j.physletb.2009.04.054.
19. A. P. Kobushkin, Ya. D. Krivenko-Emetov, S. Dubnicka. Elastic electron-deuteron scattering beyond one photon exchange. Phys. Rev. C 81 (2010) 054001.
20. D. Abbott et al. Precise Measurement of the Deuteron Elastic Structure Function $A(Q^2)$. Phys. Rev. Lett. 82 (1999) 1379. DOI: https://doi.org/10.1103/PhysRevLett.82.1379.
21. A.P. Kobushkin, Ya.D. Krivenko-Emetov, S. Dubnicka, A.Z. Dubničkova. Two-photon exchange and elastic scattering of longitudinally polarized electrons on polarized deuterons. Phys. Rev. C 84 (2011) 054007 DOI: https://doi.org/10.1103/PhysRevC.84.054007.
22. P.E. Bosted et al. Measurements of the deuteron and proton magnetic form factors at large momentum transfers Phys. Rev., C42, 38 (1990).
23. R. G. Arnold et al. Measurement of the Electron-Deuteron Elastic-Scattering Cross Section in the Range $0.8<\sim q2<\sim 6$ GeV2, Phys. Rev. Lett. 35 776 (1975).
24. L.C Alexa, et al. [JLab Hall A Collaboration] Large Momentum Transfer Measurements of the Deuteron Elastic Structure Function A(Q^2) at Jefferson Laboratory Phys.Rev.Lett.82:1374-1378,1999
25. D. Borisyuk, A. Kobushkin. Beam normal spin asymmetry of elastic eN scattering in the leading logarithm approximation Physical Review C – Nuclear Physics 73 (4), 045210